\documentclass[prb,twocolumn,floats]{revtex4}
\usepackage{graphicx,epsfig}

\begin{document}

\title{Hysteresis in the quantum Hall regimes in electron double quantum well structures}

\author{W. Pan, J.L. Reno, and J.A. Simmons}

\affiliation{Sandia National Laboratories, Albuquerque, NM 87185}

\begin{abstract}

We present in this paper experimental results on the transport hysteresis in electron double quantum well structures.
Exploring the measurement technique of fixing the magnetic field and sweeping a front gate voltage ($V_g$), we are able to
study the hysteresis by varying the top layer Landau level fillings while maintaining a relatively constant filling factor in
the bottom layer, allowing us to tackle the question of the sign of $R_{xx}$(up)-$R_{xx}$(down), where $R_{xx}$(up) is the
magnetoresistance when $V_g$ is swept up and $R_{xx}$(down) when $V_g$ swept down. Furthermore, we observe that hysteresis is
generally stronger in the even integer quantum Hall effect (IQHE) regime than in the odd-IQHE regime. This, we argue, is due
to a larger energy gap for an even-IQHE state, determined by the Landau level separation, than that for an odd-IQHE state,
determined by the Zeeman splitting.

\end{abstract}

\vskip2pc

\date{\today}
\maketitle

There is a great deal of current interest in the study of the double quantum well (DQW) structures \cite{review}. Compared to
a single layer of the two-dimensional electron or hole system (2DES or 2DHS), the existence of another layer introduces
significant interaction effects between two quantum wells. Over the years, many novel physical phenomena have been observed
\cite{eisenstein91,simmons94,manoharan97,lilly98,feng98,lok01,pillarisetty02,yusa04,eisenstein04}. In addition, since the
distance (or the coupling) between the two quantum wells can be controllably tuned from a few tenths of nanometer to several
microns, DQW structures have shown promise as possible future electronic devices for next generation information processing
\cite{simmons98}.

Recently, a new phenomenon has been discovered in the DQW structures: electronic transport hysteresis \cite{zhu00,tutuc03}.
It was observed that, when the densities of two wells are different and tunneling is negligible, the magnetotransport
coefficients show hysteretic behavior when the magnetic ($B$) field is swept up and down. This hysteretic behavior occurs
when only one QW is in the integer quantum Hall effect (IQHE) regime, and is believed to be due to a spontaneous charge
transfer between the two layers \cite{zhu00}. Specifically, when one layer enters into an IQHE state, its Fermi level jumps
from one Landau level to another. Consequently, the chemical potential between the two QW's becomes unbalanced. In reaching
an equilibrium state, a spontaneous charge transfer from one QW to the other will occur, via the ohmic contacts. Since one QW
is in the IQHE regime where the bulk is insulating, redistribution of the transferred charges takes a finite time to reach
completion. This finite time constant, combined with the finite sweeping rate of the $B$ field, gives rise to a hysteresis in
electronic transport.

This hysteretic electronic transport was first observed in a single, high electron mobility quantum well with a low mobility
parallel conducting channel \cite{zhu00}, and later in hole DQW structures \cite{tutuc03}. So far, no studies have been
conducted in the most common DQW structures, the electron DQW's. Thus, questions remain whether the hysteresis is universal
and occurs in electron DQW's.

In this paper, we present experimental results of the transport hysteresis in electron DQW structures. Exploring the
measurement technique of fixing the magnetic field and sweeping a front gate voltage ($V_g$), we are able to study the
hysteresis by varying the top layer Landau level filling while maintaining a relatively constant filling factor in the bottom
layer, allowing us to tackle the question of the sign of $R_{xx}$(up)-$R_{xx}$(down), where $R_{xx}$(up) is the
magnetoresistance when $V_g$ is swept up and $R_{xx}$(down) when $V_g$ swept down. Furthermore, we observe that hysteresis is
generally stronger in the even-IQHE regime than in the odd-IQHE regime. This, we argue, is due to a larger energy gap for an
even-IQHE state, determined by the Landau level separation, than that for an odd-IQHE state, determined by the Zeeman
splitting.

The electron DQW sample (EA1025) was MBE (molecular beam epitaxy) grown. The schematic diagram of the growth structure is
shown in Fig. 1(a). The GaAs quantum well width is 20 nm. The two QW's are separated by an Al$_{0.3}$Ga$_{0.7}$As barrier of
100nm thick. Because of this large separation, the tunneling between the two wells is negligible and the
symmetric-antisymmetric energy gap is virtually zero. Standard Hall structures with a Ti/Au Schottkey gate were fabricated.
Ohmic contacts were made by alloying Au/Ge in a forming gas at $\sim$ 420$^{\circ}$C for a few minutes. Electron transport
measurements were performed in a pumped $^3$He system with a base temperature ($T$) of $\sim 300$~mK, using the standard low
frequency ($\sim$ 13 Hz) lock-in detection techniques. The excitation current is 20 nA. Transport hysteresis was also studied
in similar DQW's of different barrier thickness. It was observed in a sample of 25nm barrier thickness. In another sample of
10 nm thickness, where the tunneling between two layers is finite, no hysteresis was observed.

\begin{figure} [h]
\centerline{\epsfig{file=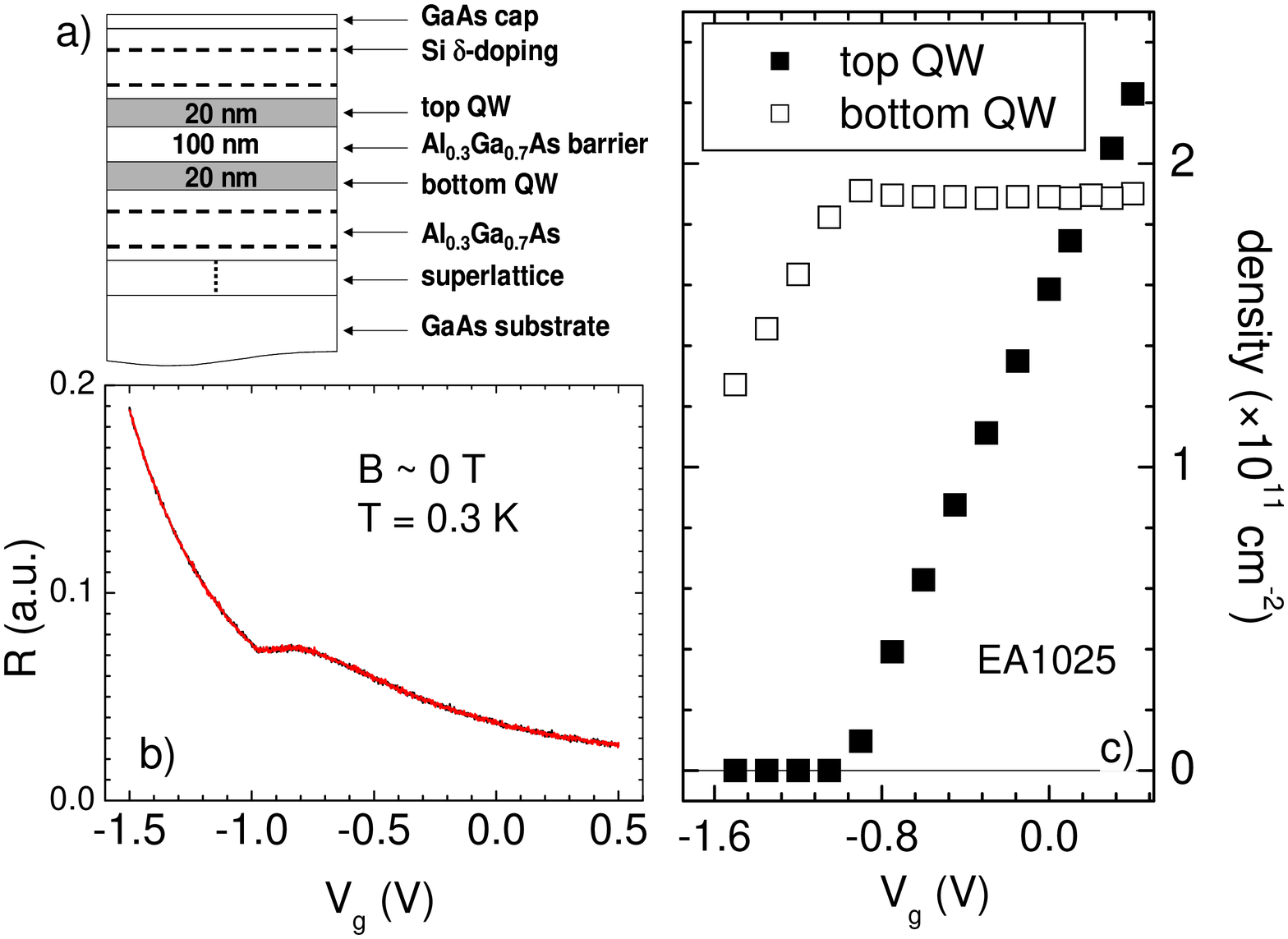,width=8.5cm}} \caption{(a) Schematic growth structure of sample EA1025 (b) Total
resistance, $R$, as a function of $V_g$. A kink is apparent when the top layer is nearly depleted. (c) Top and bottom layers
densities as a function of $V_g$. Electron densities are obtained from the FFT analysis of the low field Shubnikov-de Haas
oscillations.}
\end{figure}

Fig. 1(b) shows the results of the total resistance of two layers, $R$, as a function of $V_g$ at zero $B$ field. As $V_g$ is
negatively biased, $R$ first increases.  Close to the situation where the top layer is nearly depleted, a shallow dip shows
up. After the top layer is completely depleted, R then continuously increases as $V_g$ is further negatively biased. This
non-monotonic $V_g$ dependence was also observed in previous studies \cite{katayama95,ying95,zheng97}. In Fig. 1(c), the top
layer density ($n_{top}$) and bottom layer density ($n_{bot}$) are shown as a function of $V_g$. The densities are obtained
by performing the FFT analysis of the low-field Shubnikov-de Hass oscillations. It is clearly seen that $n_{top}$ decreases
linearly with $V_g$. From the slope of this linear dependence, a distance of $\sim$ 450 nm between the metal gate and the
center of the top layer is obtained. This value is consistent with the growth parameter of $\sim$ 410~nm. When the top layer
is totally depleted, the density of bottom layer starts to decrease. The rate of decrease is slower than that of the top
layer, consistent with a larger separation between the metal gate and the bottom layer.

Fig. 2a shows the magnetoresistance $R_{xx}$ vs. $B$ at $T = 300$~mK. These traces were obtained after illuminating the
sample with a red light emitting diode (LED). The top layer electron density is $n_{top}=2.2\times10^{11}$ cm$^{-2}$ and the
bottom layer density is $n_{bot} = 2.4\times10^{11}$ cm$^{-2}$. The total mobility is $\mu_{tot} = 2.4\times10^6$ cm$^2$/Vs.
In this slightly unbalanced DQW sample, only the even IQHE state exits \cite{boebinger90,murphy94}. Consistent with previous
studies \cite{zhu00,tutuc03}, hysteresis is observed at these IQHE states. In the temporal dependent measurements (not
shown), $R_{xx}$ in the hysteretic region shows the typical exponential decay with a time constant of ~ 1-2 minutes
\cite{zhu00}.

\begin{figure} [h]
\centerline{\epsfig{file=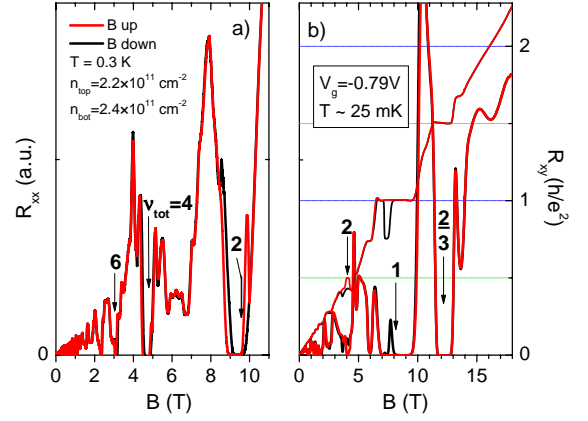,width=8.5cm}} \caption{ (a) Magnetoresistance $R_{xx}$ and Hall resistance $R_{xy}$ in
EA1025, after a brief LED illumination at 4~K. The top layer density and bottom layer density are $n_{top}=2.2\times10^{11}$
cm$^{-2}$ and $n_{bot}=2.4\times10^{11}$ cm$^{-2}$, respectively. The total mobility is $\mu_{tot} = 2.4\times10^6$
cm$^2$/vs. Hysteresis is seen at the total filling factor $\nu=2$, 4, and 6. (b) Magnetotransport coefficients in a sample of
25nm barrier thickness. $R_{xx}$ and $R_{xy}$ for $B$ sweeping up (red curve) and down (black curve) are taken at the fixed
front gate voltage of -0.79V.}
\end{figure}

Strong hysteresis is also observed when two layers are strongly imbalanced, e.g., $n_{top}/n_{bot} << 1$.  In Figure 2b, data
were taken in the DQW sample of 25nm barrier thickness at the front gate voltage of -0.79V. At this voltage, the top layer is
nearly depleted. Strong hysteresis is seen, for example, at $\nu=1$ and 2 in $R_{xx}$ as well as in the Hall resistance
$R_{xy}$. It is interesting to notice that in the $\nu=1$ hysteretic regime, $R_{xy}$ in the B sweeping down trace seems to
be quantized at a value close to 3h/4e$^2$. At the present time, it is not clear what causes this apparent quantization. On
the other hand, we note that in a recent paper Yang proposed a Wigner crystal/glass state at $\nu=1$ when the two layers are
heavily imbalanced \cite{yang01}. It remains of interest to see whether the observed Hall anomaly is related to this new
phase. When $\nu<1$, no hysteresis is seen in the FQHE regime. This is consistent with the model proposed in Ref. [12]: Once
$\nu<1$ is reached, the Fermi level will stay in the lowest Landau level and experience no more sudden jumps. Thus, no
hysteresis is expected.

In our gated samples, the magnetotransport coefficients can be measured by fixing B field while sweeping front gate voltage
($V_g$). In general, as long as the Landau level filling factor is a good quantum number, sweeping $B$ and sweeping $V_g$ (or
electron density) are equivalent. In the DQW structures, on the other hand, sweeping $V_g$ has an extra benefit. Compared to
sweeping $B$ where both the top layer filling factor ($\nu_{top}$) and the bottom layer filling factor ($\nu_{bot}$) change
simultaneously, sweeping $V_g$ allows us to vary $\nu_{top}$ alone while maintaining a relatively fixed $\nu_{bot}$. (Of
course, when charge transfers between layers, $\nu_{bot}$ changes slightly, causing the hysteresis.) In Fig. 3a, we show the
data taken at $B = 2.36$ T, or $\nu_{bot}$ =3.31 -- $R_{xx}$(up) (for $V_g$ swept from -1.5V to 0.5V) and $R_{xx}$(down) (for
$V_g$ swept from 0.5V to -1.5V). Pronounced hysteresis is observed at $\nu_{top}$ = 1, 2, 3, and 4. In Fig.3b,
$R_{xx}$(up)-$R_{xx}$(down) at various $B$ fields is plotted as a function of $V_g$. The non-zero value indicates the
occurrence of hysteresis. All the traces are shifted according to their respective $B$ field (or $\nu_{bot}$). The four
straight lines indicate the position of $\nu_{top}$ as a function of $V_g$. It is clearly seen that hysteresis occurs only
along these lines, i.e., when the top layer is in the IQHE regime.

There are a couple of new features worthwhile emphasizing in Fig. 3b. First, $R_{xx}$(up)-$R_{xx}$(down) can be either
negative or positive. As indicated in Fig. 3b, the sign depends on $\nu_{bot}$: It is positive when $\nu_{bot}$ is
$[\nu]_{bot}+\delta$, and negative when $[\nu]_{bot} = [\nu]_{bot}-\delta$, where the square brackets denote the closest
integer values to $\nu$ and $\delta < 0.5 $. Second, while hysteresis only occurs when the top layer is in the IQHE regime,
that the top layer is in the IQHE regime doesn't mean that a hysteretic electronic transport will always occur. It is also
related to $\nu_{bot}$. In Fig. 4, we plot $R_{xx}$(up) and $R_{xx}$(down) at three selective B field. At $B = 3.65$~T (or
$\nu_{bot}$ = 2.14), no hysteresis occurs in the entire gate voltage range at the experimental temperature of 0.3K. At $B =
2.36$~T (or $\nu_{bot}$=3.31), hysteresis is seen at every IQHE state. At an even smaller $B$ field, $B = 1.50$~T (or
$\nu_{bot}$ = 5.20), the situation is more interesting: Hysteresis only occurs at the even IQHE states.

\begin{figure} [h]
\centerline{\epsfig{file=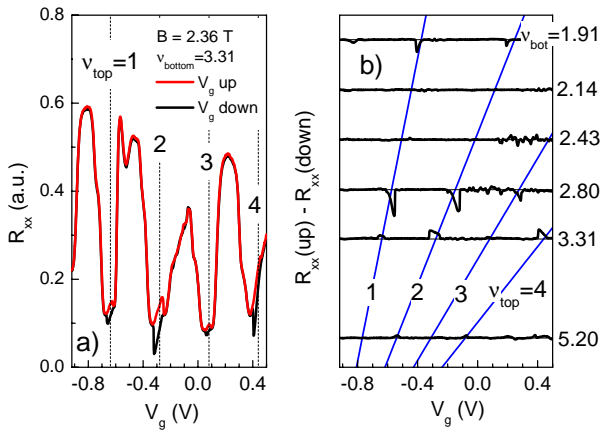,width=8.5cm}} \caption{(a) $R_{xx}$ as a function of the front gate voltage. The black
curve [$R_{xx}$(down)] is for $V_g$ sweeping down from 0.5V to -1.5V and the red curve [$R_{xx}$(up)] for $V_g$ sweeping up
from -1.5V to 0.5V. (b) $R_{xx}$(up) - $R_{xx}$(down) as a function of $V_g$. Traces are shifted vertically according to
their $B$ field values. The straight lines show the $V_g$ dependence of $\nu_{top}$ = 1, 2, 3, and 4, respectively.
$\nu_{bot}$ is also marked for each trace.}
\end{figure}

Our experimental results clearly show the transport hysteresis in the electron DQW structures. Furthermore, the hysteretic
behavior is discernable at temperatures as high as $\sim$ 600~mK, much higher than the highest temperature ($\sim$ 250~mK)
where hysteresis was previously recorded \cite{tutuc03}.  This is due to a larger electron density and a smaller electron
effective mass ($m^*$) in our electron DQW than in the hole DQW. These two factors jointly result in a larger Landau level
separation at the same $n$. Consequently, the IQHE state and hysteresis can survive at higher temperatures.

That the sign of $R_{xx}$ (up) -$R_{xx}$ (down) can be either positive or negative has also been observed in previous studies
\cite{zhu00,tutuc03} when B was varied. So far no systematic study has been conducted on this matter. In our measurements,
where $B$ is fixed and $V_g$ varied, it is apparent that at small B fields the sign shows a systematic dependence on
$\nu_{bot}$: It is positive when $\nu_{bot}= [\nu]_{bot}+\delta$ and negative when $\nu_{bot}=[\nu]_{bot}-\delta$. In the
following, we shall show that this dependence can be explained in a simple model. First, let us assume that the bottom layer
is at the Landau level filling $[\nu]_{bot}+\delta$. When $\nu_{top}$ (or $V_g$) is, for instance, decreased from
$[\nu]_{top}+\beta$ to $[\nu]_{top}$ ($\beta$ is positive and $ < 0.5$), the Fermi level jumps down. In order to reach an
equilibrium state in chemical potential between two layers, some electrons will move from the bottom QW to the top QW. In
other words, the electron density of the bottom QW decreases. Consequently, its filling factor becomes smaller and is more
close to $[\nu]_{bot}$. As a result, the resistance of the bottom QW is reduced. This, in turn, causes a reduction in
$R_{xx}$, the total resistance of the two layers. On the other hand, when $V_g$ is swept up and $\nu_{top}$ increases from
$[\nu]_{top} - \beta$ to $[\nu]_{top}$, the Fermi level jumps up. Consequently, electrons will move from the top layer to the
bottom layer. Thus, $\nu_{bot}$ increases and becomes closer to $[\nu]_{bot}$ +1/2. Since the magnetoresistance generally
displays a peak at half-fillings, the bottom layer resistance increases, resulting in an overall increase in $R_{xx}$.
Together, when $\nu_{bot} = [\nu]_{bot}+\delta$, a positive $R_{xx}$(up)-$R_{xx}$(down) is the resulting effect. The same
argument explains why the $R_{xx}$(up)-$R_{xx}$(down) is negative when $\nu_{bot} = [\nu]_{bot}-\delta$.

\begin{figure} [h]
\centerline{\epsfig{file=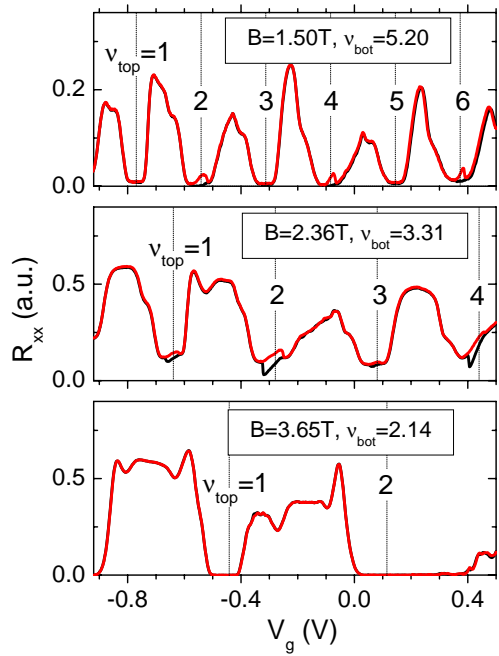,width=7.5cm}} \caption{$R_{xx}$ traces at three selective $B$ fields. The dotted lines show
the $V_g$ positions of the Landau level fillings of the top quantum well.}
\end{figure}

Another interesting observation can be made in Fig. 4: In general, hystersis is stronger in the even IQHE regime than in the
odd-IQHE regime. This seems to suggest that electron spin may also play a role. We recall that the strength of hysteresis is
related to the energy gap of an IQHE state. It is known that the energy gap of an even IQHE state is determined by the Landau
level separation, while the odd IQHE state by the Zeeman splitting. Since the effective $g$-factor for GaAs is $|g| = 0.44$,
the Landau level separation ($\hbar \omega_c = \hbar eB/m^* \sim 20\times B$[T] Kelvin) is much larger than the Zeeman
splitting ($|g|\mu_BB \sim 0.3 \times B$[T] Kelvin). This explains why in Fig. 4 the hystersis in the even IQHE regime is
stronger than that in the odd IQHE regime.

In summary, in this paper we present experimental results on transport hysteresis effects in electron double quantum well
structures. The hysteresis is studied by varying the top layer Landau level filling while maintaining a relatively constant
filling factor in the bottom layer. This measurement has allowed us to identify that the sign of $R_{xx}$(up)-$R_{xx}$(down)
is positive when $\nu_{bot}= [\nu]_{bot}+\delta$ and negative when $\nu_{bot}=[\nu]_{bot}-\delta$, where $\delta$ is a
positive number and $\delta<0.5$. A simple model is proposed to understand this sign dependence. Furthermore, it is observed
that hysteresis is generally stronger in the even-IQHE regime than in the odd-IQHE regime. This, we argue, is due to a larger
energy gap for an even-IQHE state, determined by the Landau level separation, than that for an odd-IQHE state, determined by
the Zeeman splitting.

We are grateful to Mike Lilly for lending us his 3He fridge for measurements. We thank him and Yong Chen for helpful
discussions, and E. Palm, T. Murphy, G. Jones, E. Bielejec, J. Seamons, R. Dunn, and D. Tibbetts for technical help. A
portion of this work was performed at the National High Magnetic Field Laboratory which is supported by NSF Cooperative
Agreement No. DMR-0084173 and by the State of Florida, and at the Center for Integrated Nanotechnologies, a U.S. Department
of Energy, Office of Basic Energy Sciences nanoscale science research center operated jointly by Los Alamos and Sandia
National Laboratories. Sandia National Laboratories is a multiprogram laboratory operated by Sandia Corporation, a Lockheed
Martin company, for the United States Department of Energy's National Nuclear Security Administration under contract
DE-AC04-94AL85000.


\vskip2pc

\begin{thebibliography} {90}

\bibitem{review} For a review of recent theoretical and experimental results on double quantum well structures, see, for example, the
chapters by S.M. Girvin and A.H. MacDonald, and J.P. Eisenstein in Perspectives in Quantum Hall Effect, S. Das Sarma and A.
Pinczuk (Eds.), Wiley, New York (1996), and references in.

\bibitem{eisenstein91}
J.P. Eisenstein, L.N. Pfeiffer, and K.W. West, Appl. Phys. Lett. {\bf 58}, 1497 (1991).

\bibitem{simmons94}
J.A. Simmons, S.K. Lyo, N.E. Haff, and J.F. Klem, Phys. Rev. Lett. {\bf 73}, 2256 (1994).

\bibitem{manoharan97}
H. C. Manoharan, Y. W. Suen, T. S. Lay, M. B. Santos, and M. Shayegan, Phys. Rev. Lett. {\bf 79}, 2722 (1997).

\bibitem{lilly98}
M. P. Lilly, J. P. Eisenstein, L. N. Pfeiffer, and K. W. West, Phys. Rev. Lett. {\bf 80}, 1714 (1998).

\bibitem{feng98}
X. G. Feng, S. Zelakiewicz, H. Noh, T. J. Ragucci, T. J. Gramila, L. N. Pfeiffer, and K. W. West, Phys. Rev. Lett. {\bf 81},
3219 (1998).

\bibitem{lok01}
J.G.S. Lok, S. Kraus, M. Pohlt, W. Dietsche, K. von Klitzing, W. Wegscheider, and M. Bichler, Phys. Rev. B {\bf 63}, 041305
(2001).

\bibitem{pillarisetty02}
R. Pillarisetty, H. Noh, D. C. Tsui, E. P. De Poortere, E. Tutuc, and M. Shayegan, Phys. Rev. Lett. {\bf 89}, 016805 (2002).

\bibitem{yusa04}
G. Yusa, K. Muraki, T. Saku, and Y. Hirayama, Phys. Rev. B {\bf 69}, 085323 (2004).

\bibitem{eisenstein04}
J.P. Eisenstein and A.H. MacDonald, cond-mat/0404113.

\bibitem{simmons98}
J.A. Simmons, M.A. Blount, J.S. Moon, S.K. Lyo, W.E. Baca, J.R. Wendt, J.L. Reno, and M.J. Hafich, J. Appl. Phys. {\bf 84},
5626 (1998).

\bibitem{zhu00}
J. Zhu, H.L. Stormer, L.N. Pfeiffer, K.W. Baldwin, and K.W. West, Phys. Rev. B {\bf 61}, 13361 (2000).

\bibitem{tutuc03}
E. Tutuc, R. Pillarisetty, S. Melinte, E. P. De Poortere, and M. Shayegan, Phys. Rev. B {\bf 68}, 201308 (2003).

\bibitem{katayama95}
Y. Katayama, D.C. Tsui, H.C. Manoharan, S. Parihar, and M. Shayegan, Phys. Rev. B {\bf 52}, 14817 (1995).

\bibitem{ying95}
X. Ying, S.R. Parihar, H.C. Manoharan, and M. Shayega, Phys. Rev. B {\bf 52}, 11611 (1995).

\bibitem{zheng97}
L. Zheng, M.W. Ortalano, and S. Das Sarma, Phys. Rev. B {\bf 55}, 4506 (1997).

\bibitem{boebinger90}
G.S. Boebinger, H.W. Jiang, L.N. Pfeiffer, and K.W. West, Phys. Rev. Lett. {\bf 64}, 1793 (1990).

\bibitem{murphy94}
S.Q. Murphy, J.P. Eisenstein, G.S. Boebinger, L.N. Pfeiffer, and K.W. West, Phys. Rev. Lett. {\bf 72}, 728 (1994).

\bibitem{yang01}
K. Yang, Phys. Rev. Lett. {\bf 87}, 056802 (2001).

\end{thebibliography}
\end{document}